\documentstyle[aps,prd,epsfig,preprint]{revtex}
\def\fun#1#2{\lower3.6pt\vbox{\baselineskip0pt\lineskip.9pt
\ialign{$\mathsurround=0pt#1\hfill##\hfil$\crcr#2\crcr\sim\crcr}}}
\begin{document}
\vspace{0.5in}
\title{\vskip-2.5truecm{\hfill \baselineskip 14pt 
{\hfill {{\small \hfill UT-STPD-2/99 }}}\vskip .1truecm} 
\vspace{1.0cm}
\vskip 0.1truecm {\bf Inflation, the $\mu$ Problem and Maximal 
$\nu_{\mu}-\nu_{\tau}$ Mixing}}
\vspace{1cm}
\author{{G. Lazarides}\thanks{lazaride@eng.auth.gr}} 
\vspace{1.0cm}
\address{{\it Physics Division, School of Technology, 
Aristotle University of Thessaloniki,\\ Thessaloniki GR 540 06, 
Greece}}
\maketitle

\vspace{2cm}

\begin{abstract}
\baselineskip 12pt

\par
A supersymmetric model based on a left-right symmetric 
gauge group, which `naturally' leads to hybrid inflation, 
is studied. It is shown that the $\mu$ problem can be 
easily solved in this model. The observed baryon asymmetry 
of the universe is produced via a primordial leptogenesis. 
For masses of $\nu_{\mu}$, $\nu_{\tau}$ from the small 
angle MSW resolution of the solar neutrino problem and 
SuperKamiokande, maximal $\nu_{\mu}-\nu_{\tau}$ mixing 
can be achieved. The required values of the relevant 
parameters are, however, quite small.
\end{abstract}

\thispagestyle{empty}
\newpage
\pagestyle{plain}
\setcounter{page}{1}
\def\beq{\begin{equation}}
\def\eeq{\end{equation}}
\def\beqa{\begin{eqnarray}}
\def\eeqa{\end{eqnarray}}
\def\tr{{\rm tr}}
\def\x{{\bf x}}
\def\p{{\bf p}}
\def\k{{\bf k}}
\def\z{{\bf z}}
\baselineskip 20pt

\par
A suitable framework for hybrid inflation \cite{hybrid} 
is provided \cite{lss} by a moderate extension of the 
minimal supersymmetric standard model (MSSM) based on a 
left-right symmetric gauge group. The inflaton is 
associated with the breaking of $SU(2)_{R}$ and consists 
of a gauge singlet and a pair of $SU(2)_{R}$ doublets. 
The $\mu$ problem of MSSM can be resolved \cite{dls}, 
in this scheme, by introducing \cite{lss,dls} a trilinear 
superpotential coupling of the gauge singlet inflaton 
to the electroweak higgs doublets. It has been shown 
\cite{dls} that, in the presence of gravity-mediated 
supersymmetry breaking, this gauge singlet acquires a 
vacuum expectation value (vev) and consequently generates, 
through its coupling to the electroweak higgs superfields, 
the $\mu$ term of MSSM.

\par     
The inflaton system, after the end of inflation, 
predominantly decays into electroweak higgs superfields 
and `reheats' the universe. Moreover, its subdominant 
decay into right handed neutrinos provides a mechanism 
\cite{lepto} for baryogenesis via a primordial 
leptogenesis. We solve \cite{atmo} the evolution 
equations of the inflaton system and estimate the 
`reheat' temperature. The process of baryogenesis is 
considered and its consequences on $\nu_{\mu}-
\nu_{\tau}$ mixing are analyzed \cite{atmo}. For 
masses of $\nu_{\mu}$, $\nu_{\tau}$ which are 
consistent with the small angle MSW resolution 
of the solar neutrino problem and the recent results 
of the SuperKamiokande experiment \cite{superk}, we 
examine whether maximal $\nu_{\mu}-\nu_{\tau}$ 
mixing can be achieved.

\par
Let us consider a supersymmetric model based on the
left-right symmetric gauge group $G_{LR}=
SU(3)_c\times SU(2)_L\times SU(2)_R\times 
U(1)_{B-L}$. The breaking of $SU(2)_R
\times U(1)_{B-L}$ to $U(1)_{Y}$ is achieved by 
the renormalizable superpotential 
\begin{equation}
W = \kappa S(l^c\bar l^{c}- M^2)~, 
\label{W}
\end{equation}
where $S$ is a gauge singlet chiral superfield and 
$l^c$, $\bar l^{c}$ is a conjugate pair of $SU(2)_R$ 
doublet chiral superfields which acquire superheavy 
vacuum expectation values (vevs) of magnitude $M$. The 
parameters $\kappa$ and $M$ can be made positive 
by phase redefinitions.

\par
It is well-known \cite{lss,lyth,dss} that $W$ in 
Eq.(\ref{W}) leads 
`naturally' to hybrid inflation \cite{hybrid}. This 
means that a) there is no need for `tiny' coupling 
constants, b) the superpotential is the most general 
one allowed by the symmetry, c) supersymmetry 
guarantees that radiative corrections do not invalidate 
inflation, but rather provide a slope along the 
inflationary trajectory which drives the inflaton 
towards the supersymmetric vacua, and d) supergravity 
corrections can be brought under control so as to leave 
inflation intact.

\par
The $\mu$ problem can be resolved \cite{dls} by 
introducing the extra superpotential coupling
\begin{equation}
\delta W = \lambda S  h^2 =\lambda S \epsilon^{ij}
h_i^{(1)}h_j^{(2)}~,
\label{lambda} 
\end{equation}
where the chiral electroweak higgs superfield 
$h=(h^{(1)}, h^{(2)})$ belongs to a bidoublet 
$(2,2)_{0}$ representation of $SU(2)_L\times 
SU(2)_R\times U(1)_{B-L}$ and the parameter 
$\lambda$ can again be made positive by suitable 
phase redefinitions. After gravity-mediated 
supersymmetry breaking, $S$ acquires a vev which 
generates a $\mu$ term \cite{dls}.

\par 
The scalar potential which results from the superpotential 
terms in Eqs.(\ref{W}) and (\ref{lambda}) is 
(for simplicity, we take canonical K\"ahler potential):
\begin{eqnarray}
V= | \kappa l^c\bar l^{c} + \lambda h^2 - 
\kappa M^2|^2 +(m_{3/2}^2 + \kappa^2 |\bar l^c|^2
+ \kappa^2 |l^c|^2 + \lambda^2 |h|^2)|S|^2 + 
m_{3/2}^2(|\bar l^c|^2 
\nonumber\\
+ |l^c|^2 + |h|^2) +  
\left (Am_{3/2}S( \kappa l^c\bar l^{c} + 
\lambda h^2 - \kappa M^2) + 2\kappa m_{3/2} M^{2}S+ 
{\rm h. c.} \right )~, 
\label{V}
\end{eqnarray}
where $m_{3/2}$ is the universal scalar mass (gravitino 
mass) and $A$ the universal coefficient of the trilinear
soft terms. For exact supersymmetry ($m_{3/2}
\rightarrow 0$), the vacua are \cite{dls} at
\begin{equation}
S = 0,~~~\kappa l^c\bar l^{c} + \lambda h^2 = 
\kappa M^2,~~~
l^c = e^{i\phi}\bar l^{c*}~~~h^{(1)}_i = 
e^{i\theta}\epsilon_{ij}h^{(2)j*},
\end{equation}
where the last two conditions arise from the requirement 
of D flatness. We see that there is a twofold degeneracy 
of the vacuum which is lifted by supersymmetry breaking. 
We get two degenerate (up to $m_{3/2}^{4}$) ground 
states ($\kappa\neq\lambda$): the desirable (`good') 
vacuum at $h = 0$ and $l^c\bar l^{c} = M^2$ and the 
undesirable (`bad') one at $h \neq 0$ and 
$l^c\bar l^{c} = 0$. They are separated by a potential 
barrier of order $M^{2}m_{3/2}^{2}~$.

\par
To leading order in supersymmetry breaking, the term of
the potential $V$ in Eq.(\ref{V}) proportional to $A$ 
vanishes, but a destabilizing tadpole term for $S$ 
remains:
\begin{equation} 
2\kappa m_{3/2}M^{2}S + {\rm h.c.}~.
\label{tadpole}
\end{equation} 
This term together with the mass term of $S$ 
(evaluated at the `good' vacuum) give 
$\langle S\rangle \approx - m_{3/2}/\kappa$ which, 
substituted in Eq.(\ref{lambda}), generates \cite{dls} 
a $\mu$ term with
\begin{equation}  
\mu =\lambda \langle S\rangle \approx - 
\frac{\lambda}{\kappa}m_{3/2}~.
\label{mu}
\end{equation}
Thus, coupling $S$ to the higgses can lead to the 
resolution of the $\mu$ problem.

\par
The model can be extended \cite{dls} to include matter 
fields too. The superpotential has the most general form 
respecting the $G_{LR}$ gauge symmetry and a global 
$U(1)$ R-symmetry. Baryon number is automatically 
implied by this R-symmetry to all orders in the 
superpotential, thereby guaranteeing the stability of 
proton.

\par
The model has \cite{lss,dls} a built-in inflationary 
trajectory parametrized by $|S|$, $|S| > S_c=M$ for 
$\lambda>\kappa$ (see below). All other fields vanish 
on this trajectory. The $F_S$ term is constant providing 
a constant tree level vacuum energy density 
$\kappa^2 M^4$, which is responsible for 
inflation. One-loop radiative corrections (from the mass 
splitting in the supermultiplets $l^c$, $\bar l^{c}$ 
and $h$) generate a logarithmic slope \cite{dss} along 
the inflationary trajectory which drives the inflaton 
toward the minimum. For $|S| \leq S_c=M$, the $l^c$, 
$\bar l^{c}$ components become tachyonic and the system 
evolves towards the `good' supersymmetric minimum at 
$h=0$, $l^c=\bar l^c=M$ (for $\kappa > \lambda$, $h$ 
is destabilized first and the system would have evolved 
towards the `bad' minimum at $h \neq 0$, $l^c=
\bar l^c = 0$). For all values of the parameters 
considered here, inflation continues at least till $|S|$ 
approaches the instability at $|S|=S_c$ as one deduces 
from the slow roll conditions \cite{dss,lazarides}. The 
cosmic microwave quadrupole anisotropy can be calculated 
\cite {dss} by standard methods and turns out to be   
\begin{equation}
\left(\frac{\delta T}{T}\right)_{Q} \approx 
\frac{32 \pi^{5/2}}
{3\sqrt{5}}\left(\frac{M}{M_P}\right)^{3}
\kappa^{-1}x_{Q}^{-1}\Lambda (x_{Q})^{-1}~,
\label{anisotropy}
\end{equation}
where $M_{P}=1.22\times 10^{19}~{\rm{GeV}}$ is the 
Planck scale and
\begin{eqnarray*}
\Lambda(x)=\left(\frac{\lambda}
{\kappa}\right)^{3}\left[\left(\frac{\lambda}
{\kappa}x^2-1\right)\ln \left(1-\frac{\kappa}
{\lambda}x^{-2}\right)
+\left(\frac{\lambda}{\kappa}x^2+1\right)
\ln \left(1+\frac{\kappa}
{\lambda}x^{-2}\right)\right] 
\end{eqnarray*}
\begin{equation}
+(x^2-1)\ln (1-x^{-2})+(x^2+1)\ln (1+x^{-2})~,
\label{temp}
\end{equation}
with $x=|S|/S_c$ and $S_Q$ being the value of $|S|$ 
when the present horizon scale crossed outside the 
inflationary horizon. The number of e-foldings 
experienced by the universe between the time the 
quadrupole scale exited the horizon and the end of 
inflation is 
\begin{equation} 
N_Q \approx 32 \pi^3 \left(\frac{M}{M_P}\right)^2 
\kappa^{-2}\int_{1}^{x_{Q}^{2}}\frac{dx^2}{x^2}
\Lambda(x)^{-1}~.
\label{efoldings}
\end{equation}
The spectral index of density perturbations turns out 
to be very close to unity.

\par
After reaching the instability at $|S|=S_c$, the system 
continues \cite{bl} inflating for another e-folding or 
so reducing its energy density by a factor of about $2-3$~. 
It then rapidly settles into a regular oscillatory phase 
about the vacuum. Parametric resonance is safely ignored 
in this case \cite{bl}. The inflaton (oscillating system) 
consists of the two complex scalar fields $S$ and 
$\theta=(\delta \phi + \delta\bar{\phi})/\sqrt{2}$, 
where $\delta \phi = \phi - M$, $\delta \bar{\phi} = 
\bar{\phi} - M$, with mass $m_{infl}=\sqrt{2}\kappa M$ 
($\phi$, $\bar{\phi}$ are the neutral components of 
$l^c$, $\bar l^{c}$). 

\par
The scalar fields $S$ and $\theta$ predominantly decay 
into electroweak higgsinos and higgses respectively with 
a common decay width $\Gamma_{h}=(1/16\pi)\lambda^{2}
m_{infl}$, as one can easily deduce from the couplings 
in Eqs.(\ref{W}) and (\ref{lambda}). Note, however, 
that $\theta$ can also decay to right handed neutrinos 
$\nu^c$ through the nonrenormalizable superpotential 
term
\begin{equation} 
\frac{M_{\nu^c}}{2M^{2}}\bar{\phi} 
\bar{\phi} \nu^c \nu^c~,
\label{majorana}
\end{equation} 
allowed by the gauge and R- symmetries of the model 
\cite{lss,dls}. Here, $M_{\nu^c}$ denotes the Majorana 
mass of the relevant $\nu^c$. The scalar $\theta$ decays 
preferably into the heaviest $\nu^c$ with $M_{\nu^{c}} 
\leq m_{infl}/2$. The decay rate is given by
\begin{equation}
\Gamma_{\nu^c} \approx \frac{1}{16\pi}~\kappa^2 
m_{infl}~ \alpha^2 (1-\alpha^2)^{1/2}~,
\label{decayneu}
\end{equation}
where $0\leq \alpha=2M_{\nu^c}/m_{infl} \leq 1$.
The subsequent decay of these $\nu^{c}$ 's produces 
a primordial lepton number \cite{lepto} which is then 
partially converted to the observed baryon asymmetry of 
the universe through electroweak sphaleron effects.
 
\par
The energy densities $\rho_{S}$, $\rho_{\theta}$, 
and $\rho_{r}$ of the oscillating fields $S$, $\theta$, 
and the `new' radiation produced by their decay to 
higgsinos, higgses and $\nu^c$ 's are controlled by the 
equations:
\begin{equation}
\dot{\rho}_{S}=-(3H+\Gamma_{h})\rho_{S}~,
~\rho_{\theta}(t)=\rho_{S}(t)
e^{-\Gamma_{\nu^c}(t-t_0)}~,
\label{infdensity}
\end{equation}
\begin{equation}
\dot{\rho}_{r}=-4H\rho_{r}+\Gamma_{h}\rho_{S}+
(\Gamma_{h}+\Gamma_{\nu^{c}})\rho_{\theta}~,
\label{raddensity}
\end{equation}
where
\begin{equation}
H=\frac{\sqrt{8\pi}}{\sqrt{3}M_P}~(\rho_{S}+
\rho_{\theta}+\rho_{r})^{1/2}
\label{hubble}
\end{equation}
is the Hubble parameter and overdots denote derivatives 
with respect to cosmic time $t$. The cosmic time at the 
onset of oscillations is taken $t_0\approx 0$. The 
initial values of the various energy densities are taken 
to be $\rho_{S}(t_0)=\rho_{\theta}(t_0)\approx 
\kappa^{2}M^{4}/6$, $\rho_{r}(t_{0})=0$. The `reheat' 
temperature $T_{r}$ is calculated from the equation
\begin{equation}
\rho_{S}+\rho_{\theta}=\rho_{r}=
\frac{\pi^2}{30}~g_{*}T_{r}^{4}~,
\label{reheat}
\end{equation}
where the effective number of massless degrees of 
freedom is $g_{*}$=228.75 for MSSM.

\par
The lepton number density $n_{L}$ produced by the 
$\nu^{c}$ 's satisfies the evolution equation:
\begin{equation}
\dot{n}_{L}=-3Hn_{L}+2\epsilon\Gamma_{\nu^c}
n_{\theta}~,
\label{lepton}
\end{equation}
where $\epsilon$ is the lepton number produced per 
decaying right handed neutrino and the factor of 2 in 
the second term of the rhs comes from the fact that we
get two $\nu^c$ 's for each decaying scalar $\theta$ 
particle. The `asymptotic' ($t\rightarrow 0$) lepton 
asymmetry turns out to be
\begin{equation}
\frac{n_{L}(t)}{s(t)}\sim 3
\left(\frac{15}{8}\right)^{1/4}\pi^{-1/2}
g_{*}^{-1/4}m_{infl}^{-1}~
\frac{\epsilon\Gamma_{\nu^c}}
{\Gamma_{h}+\Gamma_{\nu^c}}~\rho_{r}^{-3/4}
\rho_{S}e^{\Gamma_{h}t}~.
\label{leptonasymmetry}
\end{equation} 
For MSSM spectrum between $100$ GeV and $M$, the observed 
baryon asymmetry is then given \cite{ibanez} by 
$n_{B}/s=-(28/79)(n_{L}/s)$. It is, however, important 
to ensure that the primordial lepton asymmetry is not 
erased by lepton number violating $2\rightarrow 2$ 
scattering processes at all temperatures between $T_r$ and 
100 GeV. This requirement gives \cite{ibanez} 
$m_{\nu_{\tau}}\stackrel{_<}{_\sim} 10~{\rm{eV}}$ 
which is readily satisfied in our case (see below).

\par
Assuming hierarchical light neutrino masses, we take 
$m_{\nu_{\mu}}\approx 2.6\times 
10^{-3}~\rm{eV}$ which is the central value of the 
$\mu$-neutrino mass coming from the small 
angle MSW resolution of the solar neutrino 
problem \cite{smirnov}. The $\tau$-neutrino mass
will be restricted by the atmospheric anomaly 
\cite{superk} in the range 
$3\times 10^{-2}~\rm{eV}\stackrel{_{<}}{_{\sim }}
m_{\nu _{\tau }}\stackrel{_{<}}{_{\sim }}11\times 
10^{-2}~\rm{eV}$. Recent analysis \cite{giunti} of 
the results of the CHOOZ experiment shows that 
the oscillations of solar and atmospheric neutrinos 
decouple. We thus concentrate on the two heaviest 
families ignoring the first one. Under these 
circumstances, the lepton number generated per 
decaying $\nu^c$ is \cite{lazarides,neu}
\begin{equation}
\epsilon=\frac{1}{8\pi}~g
\left(\frac{M_3}{M_2}\right)
~\frac{{\rm c}^{2}{\rm s}^{2}\ 
\sin 2\delta \ 
(m_{3}^{D}\,^{2}-m_{2}^{D}\,^{2})^{2}}
{|\langle h^{(1)}\rangle|^{2}~(m_{3}^{D}\,^{2}\ 
{\rm s}^{2}\ +
\ m_{2}^{D}\,^{2}{\rm \ c^{2}})}~,
\label{epsilon}
\end{equation}
where $g(r)=r\ln (1 + r^{-2})$~, 
$|\langle h^{(1)}\rangle|\approx 174~\rm{GeV}$, 
${\rm c}=\cos \theta ,\ {\rm s}=\sin \theta $, 
and $\theta$ ($0\leq \theta \leq \pi /2$) and 
$\delta$ ($-\pi/2\leq \delta <\pi/2 $) are the 
rotation angle and phase which diagonalize the 
Majorana mass matrix of $\nu^{c}$ 's with eigenvalues 
$M_2$, $M_3$ ($\geq 0$). The `Dirac' mass matrix 
of the neutrinos is considered diagonal with 
eigenvalues $m_{2}^{D}$, $m_{3}^{D}$ ($\geq 0$). 

\par
For the range of parameters considered here, the 
scalar $\theta$ decays into the second heaviest right 
handed neutrino with mass $M_{2}$ ($<M_{3}$) and, 
thus, $M_{\nu^{c}}$ in Eqs.(\ref{majorana}) and 
(\ref{decayneu}) should be 
identified with $M_{2}$. Moreover, $M_{3}$ turns out 
to be bigger than $m_{infl}/2$ as it should. We will 
denote the two positive eigenvalues of the light neutrino 
mass matrix by $m_{2}$ (=$m_{\nu _{\mu }}$), $m_{3}$ 
(=$m_{\nu _{\tau }}$) with $m_{2}\leq m_{3}$. All the 
quantities here (masses, rotation angles and phases) are 
`asymptotic' (defined at the grand unification scale 
$M_{GUT}$). 

\par
The determinant and the trace invariance of 
the light neutrino mass matrix imply\cite{neu} 
two constraints on the (asymptotic) parameters which 
take the form: 
\begin{equation}
m_{2}m_{3}\ =\ \frac{\left( m_{2}^{D}m_{3}^{D}
\right) ^{2}}{M_{2}\ M_{3}}~,
\label{determinant}
\end{equation}
\begin{eqnarray*}
m_{2}\,^{2}+m_{3}\,^{2}\ =\frac{\left( m_{2}^{D}\,\,
^{2}{\rm c}^{2}+m_{3}^{D}\,^{2}{\rm s}^{2}\right) 
^{2}}{M_{2}\,^{2}}+
\end{eqnarray*}
\begin{equation}
\ \frac{\left( m_{3}^{D}\,^{2}{\rm c}^{2}+m_{2}^{D}\,
^{2}{\rm s}^{2}\right)^{2}}{M_{3}\,^{2}}+\ 
\frac{2(m_{3}^{D}\,^{2}-m_{2}^{D}\,^{2})^{2}
{\rm c}^{2}{\rm s}^{2}\,{\cos 2\delta }}
{M_{2}\,M_{3}}~\cdot
\label{trace} 
\end{equation}

\par
The $\mu-\tau$ mixing angle $\theta _{23}$ 
(=$\theta _{\mu\tau}$) lies \cite{neu} in the range
\begin{eqnarray*}
|\,\varphi -\theta ^{D}|\leq \theta _{23}\leq
\varphi +\theta ^{D},\ {\rm {for}\ \varphi +
\theta }^{D}\leq \ \pi /2~,~~~~~
\end{eqnarray*}
\begin{equation}
|\,\varphi -\theta ^{D}|\leq \theta _{23}\leq
\pi-\varphi -\theta ^{D},\ {\rm {for}\ \varphi +
\theta }^{D}\geq \ \pi /2~,
\label{mixing}
\end{equation}
where $\varphi$ ($0\leq \varphi \leq \pi /2$) is the 
rotation angle which diagonalizes the light neutrino mass 
matrix, and $\theta ^{D}$ ($0\leq \theta ^{D} \leq 
\pi /2$) is the `Dirac' (unphysical) mixing angle in 
the $2-3$ leptonic sector defined in the absence of the 
Majorana masses of the $\nu^{c}$ 's.

\par
Assuming approximate $SU(4)_{c}$ symmetry, we get the 
asymptotic (at $M_{GUT}$) relations: 
\begin{equation}
m_{2}^{D}\approx m_{c}\ ,
\ m_{3}^{D}\approx \ m_{t}\ ,
\ \sin\theta ^{D}\approx |V_{cb}|~.
\label{asympt}
\end{equation}
Renormalization effects, for MSSM spectrum and  
$\tan \beta \approx m_{t}/m_{b}$, are incorporated 
\cite{neu} by substituting in the above 
formulas the values: $m_{2}^{D}\approx 0.23~{\rm GeV}$, 
$\ m_{3}^{D}\approx 116$ GeV and $\sin \theta ^{D}
\approx 0.03$. Also, $\tan^{2} 2 \theta _{23}$ 
increases by about 40\% from $M_{GUT}$ to $M_{Z}$.

\par
We take a specific MSSM framework \cite{als} 
where the three Yukawa couplings of the third generation 
unify `asymptotically' and, thus, 
$\tan \beta \approx m_{t}/m_{b}$. We choose the 
universal scalar mass (gravitino mass) $m_{3/2} 
\approx 290~{\rm{GeV}}$ and the universal gaugino mass 
$M_{1/2} \approx 470~{\rm{GeV}}$. These values 
correspond \cite{asw} to $m_{t}(m_{t})\approx 166~
{\rm{GeV}}$ and $m_{A}$ (the tree level CP-odd scalar 
higgs mass) =$M_{Z}$. The ratio $\lambda/\kappa$ is 
evaluated \cite{carena} from 
\begin{equation}
\frac{\lambda}{\kappa}=\frac{|\mu|}{m_{3/2}}\approx 
\frac{M_{1/2}}{m_{3/2}}\left(1-\frac{Y_{t}}
{Y_{f}}\right)^{-3/7}\approx 3.95~,
\label{renorm}
\end{equation} 
where $Y_{t}=h_{t}^2\approx 0.91$ is the square of the 
top-quark Yukawa coupling and $Y_{f}\approx 1.04$ is the 
weak scale value of $Y_{t}$ corresponding to `infinite' 
value at $M_{GUT}$.

\par
\begin{figure}
\epsfig{figure=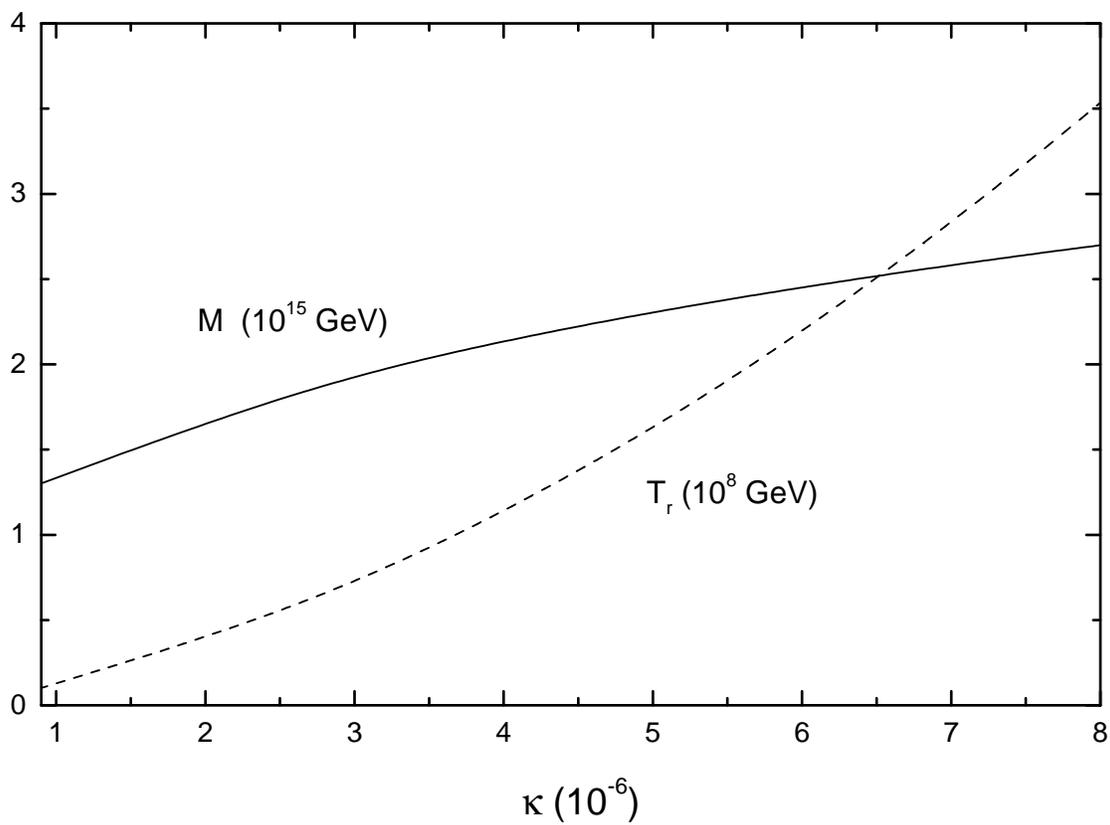,height=5.8in,angle=-90}
\medskip
\caption{The mass scale $M$ (solid line) and the reheat 
temperature $T_{r}$ (dashed line) as functions of 
$\kappa$.
\label{reheating}}
\end{figure}

Eqs.(\ref{anisotropy})-(\ref{efoldings}) can now be 
solved, for $(\delta T/T)_{Q} \approx 6.6\times 10^{-6}$ 
from COBE, $N_Q \approx 50$ and any value of $x_{Q}>1$. 
Eliminating $x_{Q}$, we obtain $M$ as a function of 
$\kappa$ depicted in Fig.\ref{reheating}. The evolution 
Eqs.(\ref{infdensity})-(\ref{hubble}) are solved 
for each value of $\kappa$. The parameter $\alpha^{2}$ 
in Eq.(\ref{decayneu}) is taken equal to 2/3. This 
choice maximizes the decay width of the inflaton 
to $\nu^{c}$ 's and, thus, the subsequently produced 
lepton asymmetry. The `reheat' temperature is then 
calculated from Eq.(\ref{reheat}) for each value of 
$\kappa$. The result is again depicted in 
Fig.\ref{reheating}.

\par
\begin{figure}
\epsfig{figure=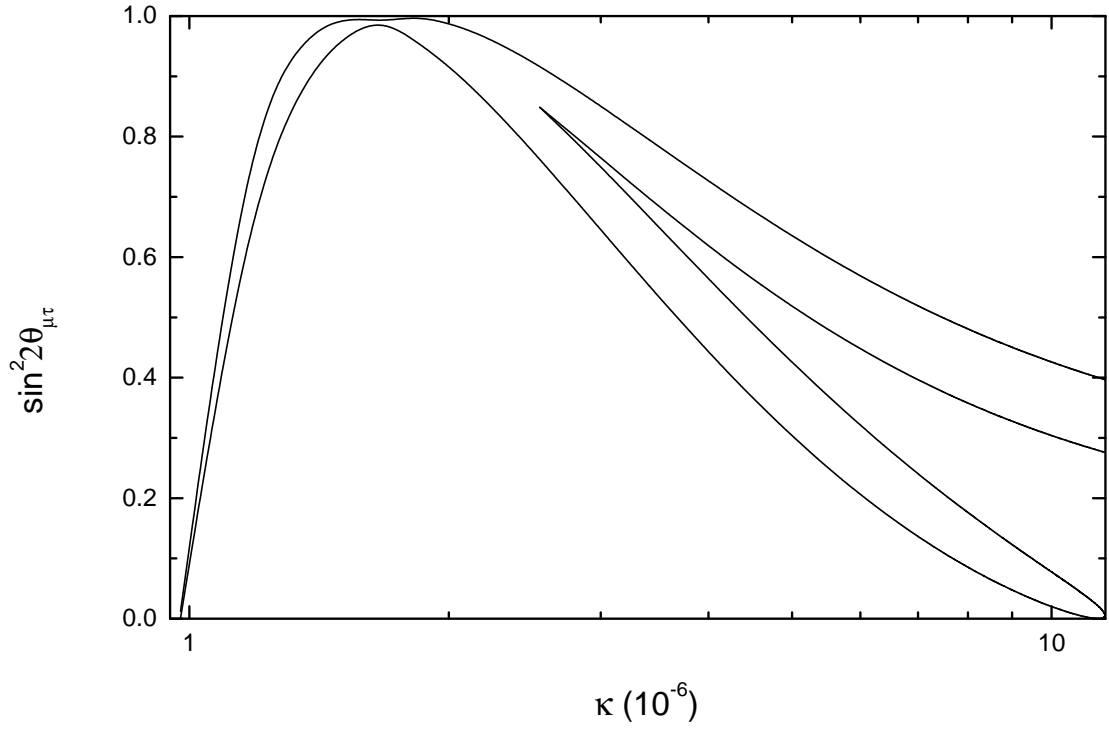,height=5.8in,angle=-90}
\medskip
\caption{The allowed region (bounded by the solid lines)
in the $\kappa-\sin^{2} 2 \theta _{\mu\tau}$ plane 
for $m_{\nu_{\mu}}\approx 2.6\times 10^{-3}~\rm{eV}$ 
and $m_{\nu_{\tau}}\approx 7 \times 10^{-2}~\rm{eV}$.
\label{angle}}
\end{figure}

We next evaluate the lepton asymmetry. We first take 
$m_{\nu_{\tau}}\approx 7 \times 10^{-2}~\rm{eV}$, 
the central value  from SuperKamiokande \cite{superk}. 
The mass of the second heaviest $\nu^c$, into which the 
scalar $\theta$ decays partially, is given by $M_{2}=
M_{\nu^{c}}=\alpha m_{infl}/2$ and $M_{3}$ is found 
from the `determinant' condition in Eq.(\ref{determinant}). 
The `trace' condition in Eq.(\ref{trace}) is then solved 
for $\delta(\theta )$ which is subsequently substituted 
in Eq.(\ref{epsilon}) for $\epsilon$. The leptonic 
asymmetry as a function of the angle $\theta$ can be found 
from Eq.(\ref{leptonasymmetry}). For each value of 
$\kappa$, there are two values of $\theta$ satisfying the 
low deuterium abundance constraint $\Omega _{B}h^{2}
\approx 0.025$. (These values of $\theta$ turn out to be 
quite insensitive to the exact value of $n_{B}/s$.) The 
corresponding $\varphi$ 's are then found and the allowed 
region of the mixing angle $\theta _{\mu\tau}$ 
in Eq.(\ref{mixing}) is determined for each $\kappa$. 
Taking into account renomalization effects and 
superimposing all the permitted regions, we obtain the 
allowed range of $\sin^{2} 2 \theta _{\mu\tau}$ 
as a function of $\kappa$, shown in Fig.\ref{angle}. 
We observe that maximal mixing
($\sin^{2} 2 \theta _{\mu\tau}\approx 1$) is achieved 
for $1.5\times 10^{-6}\stackrel{_{<}}{_{\sim }}\kappa
\stackrel{_{<}}{_{\sim }}1.8\times 10^{-6}$. Also, 
$\sin^{2} 2 \theta _{\mu\tau}\stackrel{_{>}}
{_{\sim }} 0.8$ \cite{superk} corresponds to 
$1.2\times 10^{-6}\stackrel{_{<}}{_{\sim }}\kappa
\stackrel{_{<}}{_{\sim }}3.4\times 10^{-6}$. 

\par
\begin{figure}
\epsfig{figure=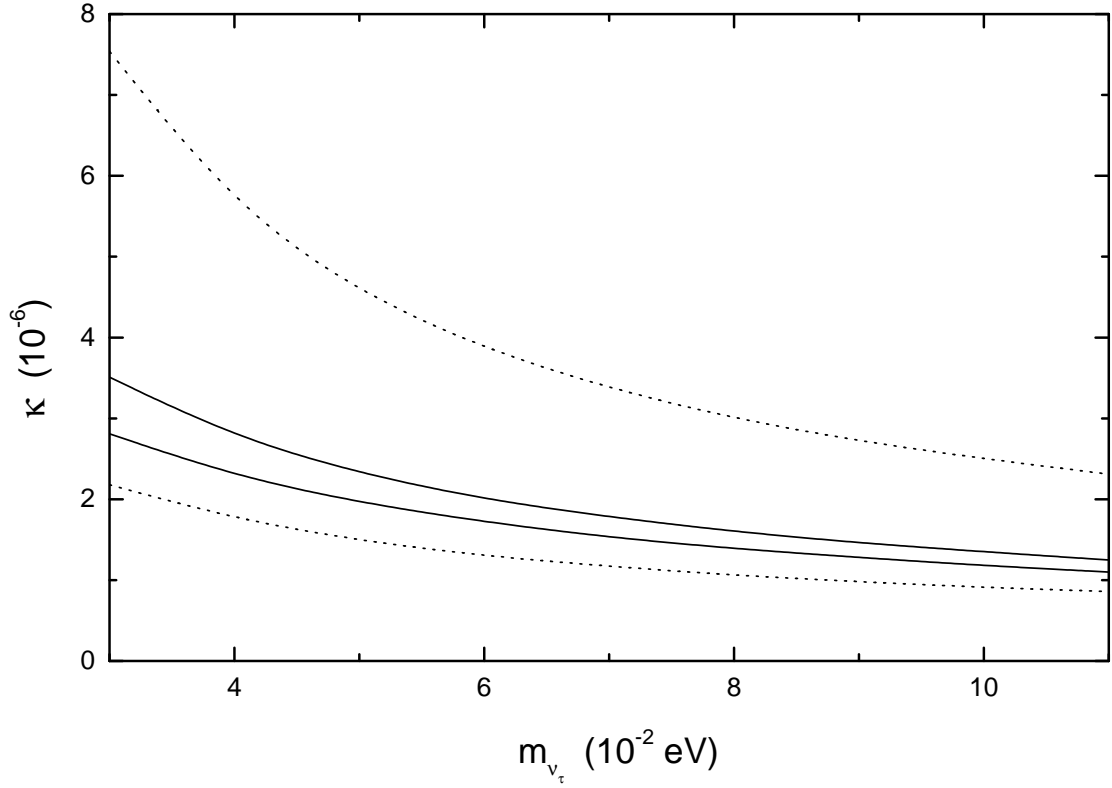,height=5.8in,angle=-90}
\medskip
\caption{The regions on the $m_{\nu_{\tau}}-\kappa$
plane corresponding to maximal $\nu_{\mu}-\nu_{\tau}$ 
mixing (bounded by the solid lines) and 
$\sin^{2} 2 \theta _{\mu\tau}
\stackrel{_{>}}{_{\sim }} 0.8$ (bounded by the dotted 
lines). Here we consider the range $3\times 
10^{-2}~\rm{eV}\stackrel{_{<}}{_{\sim }}
m_{\nu _{\tau }}\stackrel{_{<}}{_{\sim }}
11\times 10^{-2}~\rm{eV}$ ($m_{\nu_{\mu}}
\approx 2.6\times 10^{-3}~\rm{eV}$).
\label{kappa}}
\end{figure}

We repeated the above analysis for all values of
$m_{\nu _{\tau }}$ allowed by SuperKamiokande. 
The allowed regions in the $m_{\nu _{\tau }}-\kappa$ 
plane for maximal $\nu_{\mu}-\nu_{\tau}$ mixing 
(bounded by the solid lines) and $\sin^{2} 2 
\theta _{\mu\tau}\stackrel{_{>}}{_{\sim }} 0.8$ 
(bounded by the dotted lines) are shown in 
Fig.\ref{kappa}. Notice that, for $\sin^{2} 2 
\theta _{\mu\tau}\stackrel{_{>}}{_{\sim }} 0.8$, 
$\kappa\approx(0.9-7.5)\times 10^{-6}$ which is 
rather small. (Fortunately, supersymmetry protects it 
from radiative corrections.) The corresponding values of 
$M$ and $T_{r}$ can be read from Fig.\ref{reheating}. 
We find $1.3\times 10^{15}~{\rm{GeV}}
\stackrel{_{<}}{_{\sim }} M
\stackrel{_{<}}{_{\sim }}2.7\times 10^{15}~{\rm{GeV}}$ 
and $10^{7}~{\rm{GeV}}\stackrel{_{<}}{_{\sim }} T_{r}
\stackrel{_{<}}{_{\sim }}3.2\times 10^{8}~{\rm{GeV}}$. 
We observe that $M$ turns out to be somewhat smaller than 
the MSSM unification scale $M_{GUT}$. (It is anticipated 
that $G_{LR}$ is embedded in a grand unified theory.) The 
reheat temperature, however, satisfies the gravitino 
constraint 
($T_{r}\stackrel{_{<}}{_{\sim}} 10^{9}~{\rm{GeV}}$).
Note that, for the values of the parameters
chosen here, the lightest supersymmetric particle (LSP)
is \cite{asw} an almost pure bino with mass $m_{LSP}
\approx 0.43M_{1/2}\approx 200~{\rm{GeV}}$ \cite{drees}
and can, in principle, provide the cold dark matter of the 
universe. On the contrary, there is no hot dark matter 
candidate, in the simplest scheme.

\par
In conclusion, we have shown that, in a supersymmetric 
model based on a left-right symmetric gauge group and 
leading `naturally' to hybrid inflation, the $\mu$ 
problem can be easily solved . The observed baryon 
asymmetry of the universe is produced via a primordial 
leptogenesis. For masses of $\nu_{\mu}$, $\nu_{\tau}$ 
from the small angle MSW resolution of the solar neutrino 
puzzle and SuperKamiokande, maximal 
$\nu_{\mu}-\nu_{\tau}$ mixing can be achieved. The 
required values of the coupling constant $\kappa$ are, 
however, quite small ($\sim 10^{-6}$).
 
\vspace{0.5cm}
This work is supported by E.U. under TMR contract 
No. ERBFMRX--CT96--0090.

\def\ijmp#1#2#3{{ Int. Jour. Mod. Phys. }
{\bf #1~}(19#2)~#3}
\def\pl#1#2#3{{ Phys. Lett. }{\bf B#1~}(19#2)~#3}
\def\zp#1#2#3{{ Z. Phys. }{\bf C#1~}(19#2)~#3}
\def\prl#1#2#3{{ Phys. Rev. Lett. }{\bf #1~}(19#2)~#3}
\def\rmp#1#2#3{{ Rev. Mod. Phys. }{\bf #1~}(19#2)~#3}
\def\prep#1#2#3{{ Phys. Rep. }{\bf #1~}(19#2)~#3}
\def\pr#1#2#3{{ Phys. Rev. }{\bf D#1~}(19#2)~#3}
\def\np#1#2#3{{ Nucl. Phys. }{\bf B#1~}(19#2)~#3}
\def\mpl#1#2#3{{ Mod. Phys. Lett. }{\bf #1~}(19#2)~#3}
\def\arnps#1#2#3{{ Annu. Rev. Nucl. Part. Sci. }{\bf
#1~}(19#2)~#3}
\def\sjnp#1#2#3{{ Sov. J. Nucl. Phys. }{\bf #1~}
(19#2)~#3}
\def\jetp#1#2#3{{ JETP Lett. }{\bf #1~}(19#2)~#3}
\def\app#1#2#3{{ Acta Phys. Polon. }{\bf #1~}(19#2)~#3}
\def\rnc#1#2#3{{ Riv. Nuovo Cim. }{\bf #1~}(19#2)~#3}
\def\ap#1#2#3{{ Ann. Phys. }{\bf #1~}(19#2)~#3}
\def\ptp#1#2#3{{ Prog. Theor. Phys. }{\bf #1~}
(19#2)~#3}
\def\plb#1#2#3{{ Phys. Lett. }{\bf#1B~}(19#2)~#3}

\newpage

\pagestyle{empty}

\end{document}